\documentstyle[12pt,epsfig]{article}
\def\e{\epsilon}

\def\ycut{y_{\rm cut}}

\def\smin{s_{\rm min}}
\begin{document}
\pagestyle{plain}

\begin{titlepage}
\vspace*{-1cm}
\begin{flushright}
June 1998 \\
\end{flushright}                                
\vskip 1.cm
\renewcommand{\thefootnote}{\fnsymbol{footnote}}
\begin{center}                                                             
{\Large\bf
Isolated and non-isolated photon rates at LEP
\footnote{Talk presented at the Zeuthen Workshop on Elementary Particle Theory
``Loops and Legs in Gauge Theories" Rheinsberg, Germany, April 19-24, 1998.}}
\vskip 1.3cm

{\large A.~Gehrmann--De Ridder}
\vskip .2cm
{\it DESY, Theory Group, D-22603 Hamburg, Germany}
\vskip 2.3cm   
\end{center}      
\begin{abstract}
We present the results of the calculation of the `photon' +~1 jet rate 
at ${\cal O}(\alpha \alpha_{s})$ for LEP energies.
By comparing these results with the data from the ALEPH Collaboration
we make a next-to-leading order determination of the quark-to-photon 
fragmentation function $D_{q\to \gamma}(z,\mu_{F})$ at 
${\cal O}(\alpha \alpha_{s})$. The predictions obtained using this 
fragmentation function for the {\it isolated} rate, defined as the  
`photon' +~1 jet rate for $z>0.95$, are found in good agreement with 
the ALEPH data. The next-to-leading order corrections are moderate 
demonstrating the perturbative stability of this particular 
isolated photon definition.
We have also computed the inclusive photon energy 
distribution and found good agreement with the OPAL data.  
\end{abstract}                                                                
\vfill
\end{titlepage}
\newpage
\renewcommand{\thefootnote}{\arabic{footnote}} 
\setcounter{footnote}{0}
\section{Introduction}
Photons produced in hadronic collisions arise essentially 
from two different sources: the {\em direct} production of a 
photon off a primary parton or  
through the {\em fragmentation} of a hadronic jet into a single photon 
carrying a large fraction of the jet energy.
The former gives rise to perturbatively calculable short-distance 
contributions whereas the latter is primarily a long 
distance process which cannot be calculated within perturbative QCD. 
It is described by the process-independent parton-to-photon 
fragmentation function~\cite{phofrag} 
which must be determined from experimental data. 
Its evolution with the factorization scale $\mu_{F}$ can however 
be determined by perturbative methods.

The ALEPH Collaboration at CERN has measured
the quark-to-photon fragmentation function 
$D_{q \to \gamma}$~\cite{aleph} 
from an analysis of two jet events in which one of the jets contains a 
photon carrying a large fraction ($z>0.7$) of the jet energy. 
These `photon' +1~jet events are defined by the application of the 
Durham jet clustering algorithm \cite{durham} to both the hadronic 
and electromagnetic clusters. In this democratic approach, the photon is 
called {\em isolated} if it carries a large fraction, 
typically 95\%, of the jet energy and said to be non-isolated otherwise. 
A comparison between this measured rate and the calculated rate 
up to ${\cal O}(\alpha)$ \cite{andrew} using the same 
clustering approach to define the photon 
yielded a first determination \cite{aleph} 
of the quark-to-photon fragmentation function 
accurate at this order. Furthermore, the insertion of this measured function 
into the ${\cal O}(\alpha)$ `isolated' rate, defined as the `photon' +1~jet
rate for $z>0.95$, yielded a good agreement with the ALEPH data.  

In this fixed order framework,  
the distribution of electromagnetic energy within 
the photon jet of photon + 1 jet events, for a single quark of 
charge $e_q$, at ${\cal O}(\alpha)$ can be written \cite{andrew},
\begin{equation}
\frac{1}{\sigma_0} \frac{d\sigma}{dz}
= 2 D_{q\to \gamma}(z,\mu_F) + \frac{\alpha e_q^2}{\pi} 
P_{q \gamma}^{(0)}(z) \log \left(\frac{s_{\rm cut}}{\mu_F^2}\right) \,+\,
R_{\Delta}\delta(1-z) + \ldots,
\label{eq:sig0}
\end{equation}
where $\ldots$ represents terms which are 
well behaved as $z \to 1$.
In the Durham jet algorithm and at large $z$, $s_{\rm cut} 
\sim sz(1-z)^2/(1+z) \sim p_T^2$ \cite{andrew} where $p_T$ is the 
transverse momentum of the photon with respect to the cluster.
The non-perturbative fragmentation function 
is an exact solution at ${\cal O}(\alpha)$ of 
the evolution equation in the factorization scale $\mu_{F}$,
\begin{equation}
D_{q\to \gamma}(z,\mu_{F}) = 
\frac{\alpha e_q^2}{2 \pi} P_{q \gamma}^{(0)}(z)
\log\left(\frac{\mu_{F}^2}{\mu_{0}^2}\right) + D_{q\to \gamma}(z,\mu_{0}).
\end{equation}
In this equation, all unknown long-distance effects are related 
to the behaviour 
of $D_{q\to \gamma}(z,\mu_{0})$, the initial value of 
this fragmentation function which has been fitted to the data 
at some initial scale $\mu_{0}$ in~\cite{aleph}.
As $D_{q\to \gamma}(z,\mu_{F})$ is exact, 
this solution does not take the commonly implemented 
\cite{grv} resummations of $\log(\mu_{F}^2)$ into account and when 
used to evaluate the photon +1 jet rate at ${\cal O}(\alpha)$ yields a 
factorization scale independent prediction for the cross section. 

In the conventional approach, a resummation of the logarithms 
of the factorization scale is performed to all orders in $\alpha_{s}$
\cite{grv} and the solution of the evolution equation for 
$D_{q\to \gamma}$ is proportional to $\log\left(\mu_{F}^2/\mu_{0}^2 \right)$. 
For $z<1$, we find that $\mu_F^2 \sim s_{\rm cut}$   
and $\mu_F^2 \gg \mu_{0}^2$ . The `direct' contribution 
in eq.~(\ref{eq:sig0}) is therefore suppressed relative to the fragmentation contribution. 
The conventional assignment of a power 
of $1/\alpha_s$ to the fragmentation function is in this case 
clearly motivated, this contribution is indeed more significant. 
However, as $z\to 1$, 
we see that the transverse size of the photon jet cluster 
decreases such that $s_{\rm cut}\to 0$.
The hierarchy $s_{\rm cut} \sim \mu_F^2$ and $\mu_F^2 \gg \mu_0^2$ is
no longer preserved and both contributions in
eq.~(\ref{eq:sig0}) are important.
Large logarithms of $(1-z)$ become the most dominant contributions.
Being primarily interested in the high $z$ region, in \cite{andrew}
it was chosen not to impose the conventional prejudice 
 and 
resum the logarithms of $\mu_F$ {\em a priori} but to work 
within a fixed order framework, to isolate the relevant large logarithms.  
A detailed comparison of the evaluation of the photon production
cross section in the conventional and fixed order formalisms can be found in 
\cite{papernew}.

We have performed the calculation of the
${\cal O}(\alpha_s)$ corrections to the 
`photon' +~1 jet rate using the same democratic procedure 
to define the photon as in~\cite{aleph,andrew}.
The details of this fixed order calculation 
have been presented in ~\cite{big}. We shall here 
limit ourselves to outline the main characteristics of this calculation, 
to summarize the results and 
to show how these compare with the available experimental data.

\section{The calculation of the `photon' +~1 jet rate 
at ${\cal O}(\alpha \alpha_s)$}
\begin{figure}[t]
\begin{center}
~ \epsfig{file=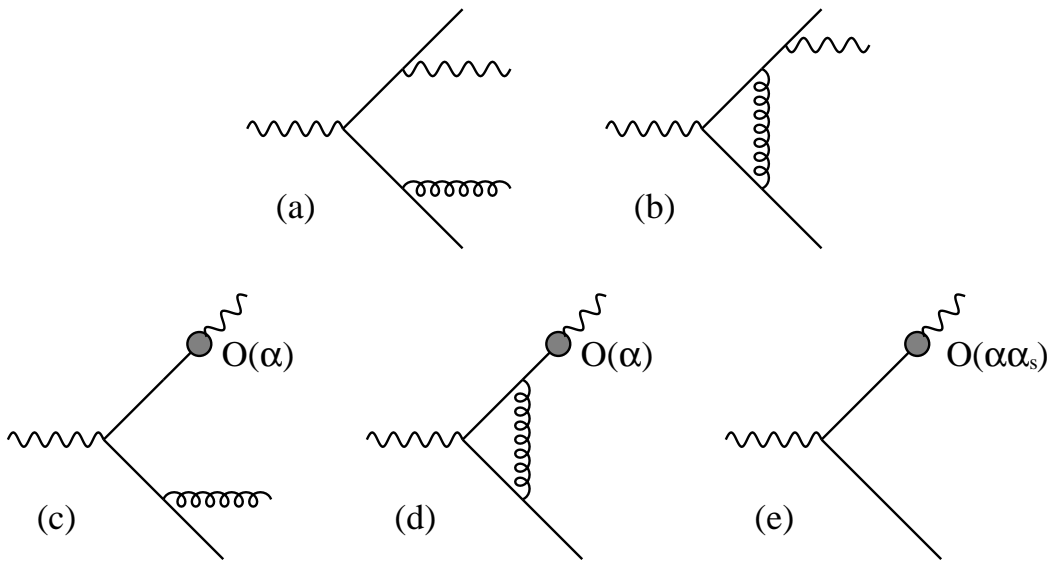,width=14cm}
\caption{Parton level subprocesses contributing to the photon +~1 jet 
rate at ${\cal O}(\alpha\alpha_s)$.} 
\label{fig:class}
\end{center}
\end{figure}
The `photon' +~1 jet rate in $e^+e^-$ annihilation at ${\cal O} (\alpha 
\alpha_s)$ receives contributions from five parton-level subprocesses
displayed in Fig.~\ref{fig:class}.
Although the `photon' +~1 jet cross section 
is finite at ${\cal O}(\alpha \alpha_{s})$, 
all these contributions  
contain divergences (when the photon 
and/or the gluon are collinear with one of the quarks, 
when the gluon is soft or since the bare quark-to-photon fragmentation 
function contains infinite counter terms).
All these divergences have to be isolated and cancelled analytically
before the `photon' +~1 jet cross section can be evaluated numerically.

Within each singular region which we have defined 
using a theoretical criterion $\smin$, 
the matrix elements are approximated and 
the unresolved variables analytically integrated.
The evaluation of the singular contributions 
associated with the process $\gamma ^* \to q \bar{q}g \gamma$ is of 
particular interest as it contains various ingredients 
which could directly be applied to the calculation of jet observables 
at next-to-next-to-leading order.
Indeed, besides the contributions arising when 
one final state gluon is collinear or soft, 
there are also contributions where {\em two} of the 
final state partons are theoretically unresolved.
The three different double unresolved contributions which 
occur in this calculation are: 
the {\it triple collinear} contributions, 
arising when the photon and the gluon are simultaneously collinear 
to one of the quarks, the {\it soft/collinear} contributions 
arising when the photon is collinear to one of the quarks 
while the gluon is soft and the {\it double single collinear} contributions, 
resulting when the photon is collinear to one of the quarks while the gluon 
is collinear to the other.
A detailed derivation of each of these singular real contributions 
and of the singular contributions arising in the processes depicted 
in Fig.~\ref{fig:class}(b)-(d) has been presented in~\cite{big}. 

Combining all unresolved contributions present in the 
processes shown in Fig.~\ref{fig:class}(a)-(d)  
yields a result 
that still contains single and double poles in $\e$.
These pole terms are however proportional 
to the universal next-to-leading order splitting function 
$P_{q\gamma}^{(1)}$ ~\cite{curci} and a convolution of two 
lowest order splitting functions, 
$(P_{qq}^{(0)}\otimes P_{q \gamma}^{(0)})$. 
Hence, they can be factorized into the next-to-leading order 
counterterm of the bare quark-to photon fragmentation function 
\cite{fac} present in the contribution depicted in Fig.~\ref{fig:class}(e),
yielding a finite and factorization scale ($\mu_{F}$) dependent 
result \cite{big}.

We have then chosen to evaluate the remaining 
finite contributions numerically using the {\it hybrid subtraction} method, 
a generalization of the {\it phase space slicing} procedure \cite{gg,kramer}.  
This latter procedure turns out to be inappropriate   
when more than one particle is potentially unresolved.
Indeed, in our calculation we found areas in the four parton phase space 
which belong simultaneously to two different single collinear regions.
Those areas cannot be treated correctly within the phase space 
slicing procedure.
Within the {\it hybrid subtraction} method developed in \cite{eec}, 
a parton resolution criterion $\smin$ 
is used to separate the phase space into different resolved and 
unresolved regions, but, rather than assuming that 
the approximated matrix elements are exact in the singular regions,  
the difference between the full matrix 
element and its approximation is evaluated numerically 
in all unresolved regions. The non-singular contributions 
are calculated using Monte Carlo methods like within the phase space 
slicing scheme.
 
The numerical program finally evaluating the `photon' +~1 jet rate 
at ${\cal O}(\alpha\alpha_s)$ contains four separate contributions.
\begin{figure}[t]
\begin{center}
~ \epsfig{file=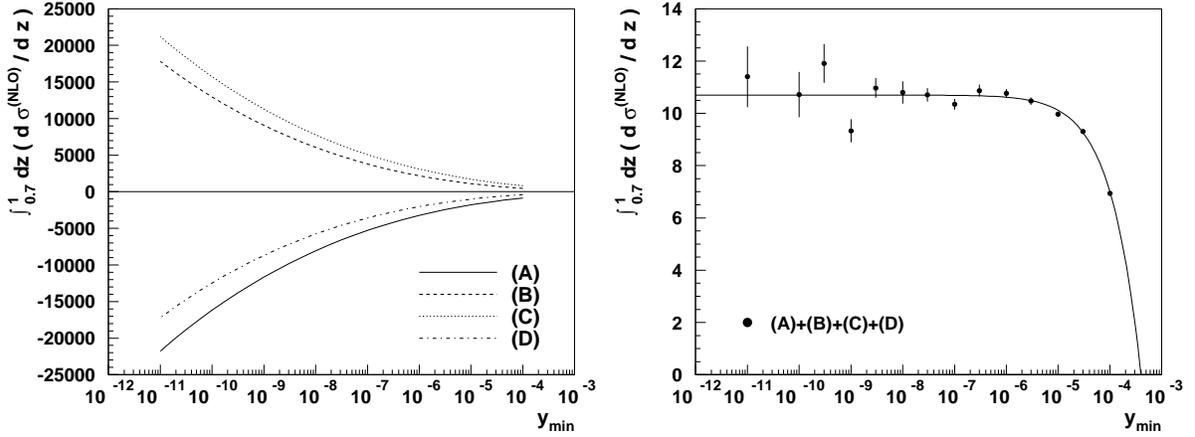,angle=-90,width=16cm}
\caption{${\cal O}(\alpha\alpha_s)$ individual contributions (left) 
and sum of all ${\cal O}(\alpha\alpha_s)$ contributions (right) to the 
photon + 1 jet rate for a single quark of charge $e_q$ such that
$\alpha e_q^2 = 2\pi$, $\alpha_s (N^2-1)/2N = 2\pi$ using
the Durham  jet algorithm with $y_{{\rm cut}}=0.1$,  
and integrated for $z>0.7$.}
\label{fig:ymin2}
\end{center}
\end{figure}
Each of them depends logarithmically (in fact 
as $\log^3 (y_{{\rm min}})$) on 
the theoretical resolution parameter $y_{{\rm min}}=s_{{\rm min}}/Q^2$. 
The physical `photon' + 1~jet cross section,
which is the sum of all four contributions,
{\it must} of course be independent of the choice of $y_{{\rm min}}$.
In Fig.~\ref{fig:ymin2}, we see that the cross section approaches (within 
numerical errors) a constant value provided that $y_{{\rm min}}$ 
is chosen small enough, indicating 
that a complete cancellation of all powers of $\log (y_{{\rm min}})$ 
takes place. This 
provides a strong check on the correctness of our results 
and on the consistency of our approach. 

\section{Results}
A comparison between the measured `photon' +~1 jet rate  \cite{aleph}
and our calculation yielded a first determination of 
the quark-to-photon fragmentation function accurate up to 
${\cal O}(\alpha \alpha_s)$ \cite{letter}.
This function, which parameterizes the 
perturbatively incalculable long-distance effects, 
has to satisfy a perturbative evolution equation in the factorization 
scale $\mu_F$. 
Indeed, the next-to-leading order fragmentation
function can be expressed as an {\it exact} solution of the evolution equation 
up to ${\cal O}(\alpha \alpha_s)$ \cite{big},
\begin{eqnarray}
D(z,\mu_{F})&=&
\frac{\alpha e_{q}^2}{2\pi}P^{(0)}_{q \gamma}(z)
\log\left(\frac{\mu^2_{F}}{\mu_{0}^2}\right) 
+\frac{\alpha e_{q}^2}{2\pi} \frac{\alpha_{s}}{2 \pi}
\left(\frac{N^2 -1}{2N}\right)P_{q \gamma}^{(1)}(z)
\log \left(\frac{\mu^2_{F}}{\mu_{0}^2}\right)
\nonumber\\
& & +
\frac{\alpha_{s}}{2 \pi}
\left(\frac{N^2 -1}{2N}\right)
\log \left(\frac{\mu^2_{F}}{\mu_{0}^2}\right) P_{qq}^{(0)}(z)\otimes 
\frac{\alpha e_{q}^2}{2 \pi}\frac{1}{2}P_{q \gamma}^{(0)}(z)
\log \left(\frac{\mu^2_{F}}{\mu_{0}^2}\right)
\nonumber\\
& &
+\frac{\alpha_{s}}{2 \pi}
\left(\frac{N^2 -1}{2N}\right)
\log \left(\frac{\mu^2_{F}}{\mu_{0}^2}\right) P_{qq}^{(0)}(z)\otimes 
D(z,\mu_{0}) \,+D(z,\mu_{0}).
\label{eq:Dnlo}
\end{eqnarray}
The initial function $D(z,\mu_{0})$ has been fitted to the ALEPH  ~1 jet
data \cite{letter} for  
$\frac{1}{\sigma_{0}}\frac{d\sigma}{dz}$,  
for the jet resolution parameter $y_{{\rm cut}}=0.06$ 
and $\alpha_s(M_z^2) = 0.124$
to yield \footnote{Note that the logarithmic term proportional to 
$P^{(0)}_{q \gamma}(z)$
is introduced to ensure that the predicted $z$ distribution is 
well behaved as $z \to 1$ \cite{andrew}.}, 
\begin{equation}
D^{NLO}(z,\mu_{0})=\frac{\alpha e_{q}^2}{2 \pi} 
\left(-P^{(0)}_{q \gamma}(z)\log(1-z)^2 \;+\,20.8\,(1-z)-11.07\right),
\label{eq:fitnlo}
\end{equation}
where $\mu_{0}=0.64$~GeV.
The next-to-leading order ($\overline{{\rm MS}}$)
quark-to-photon fragmentation function (for a quark of unit charge) 
at a factorization scale $\mu_F=M_Z$ were shown 
in \cite{letter} and 
compared with the lowest order fragmentation function obtained 
in~\cite{aleph}. A large difference between the leading and next-to-leading 
order quark-to-photon fragmentation functions was observed only for  
$z$ close to 1, indicating the presence of large $\log (1-z)$. 

Moreover, a comparison between 
the ALEPH data and the results of the ${\cal O}(\alpha \alpha_s)$ 
calculation using the fitted next-to-leading order 
fragmentation function for different values of $\ycut$ 
can be found in \cite{big,letter}.
The next-to-leading order corrections were found to be moderate 
for all values of $\ycut$ demonstrating the 
perturbative stability of our fixed order approach.
\begin{figure}[t]
\begin{center}
~ \epsfig{file=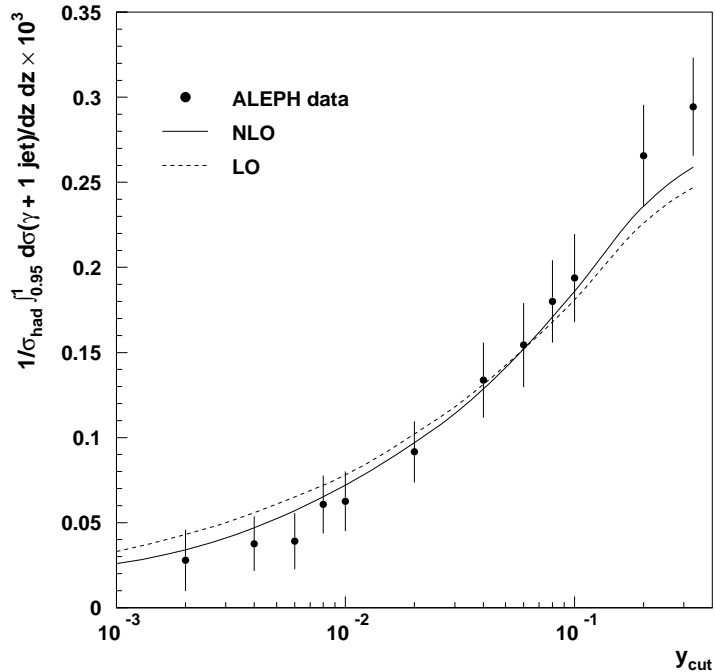,width=9cm,angle=-90}
\caption{The integrated photon +~1 jet rate above $z=0.95$ as function of
$y_{{\rm cut}}$, compared with the ${\cal O}(\alpha)$ 
and ${\cal O}(\alpha \alpha_s)$
order calculations including the appropriate 
quark-to-photon fragmentation functions.}
\label{fig:ycut}
\end{center}
\end{figure}
To test the generality of our results, we have considered 
two further applications: 
the `isolated' photon rate and the inclusive photon distribution 
which we shall now briefly present.
  
Using the results of the calculation of the photon +~1 jet rate at 
${\cal O}(\alpha \alpha_s)$ and  
the fitted quark-to-photon fragmentation function, we have 
determined the {\it isolated} 
rate defined as the photon +~1 jet rate for $z>0.95$ 
in the democratic approach. 
The result of this calculation compared with 
data from ALEPH~\cite{aleph} and the leading order calculation~\cite{andrew}
is shown in Fig.~\ref{fig:ycut}. It can clearly be seen that inclusion of 
the next-to-leading order corrections improves the agreement between 
data and theory over the whole range of $y_{{\rm cut}}$. 
It is also apparent that the next-to-leading order corrections 
to the isolated photon +~1 jet rate obtained in this 
democratic clustering approach are of reasonable size 
indicating a good perturbative stability of this {\it isolated} 
photon definition. 
\begin{figure}[t]
\begin{center}
~ \epsfig{file=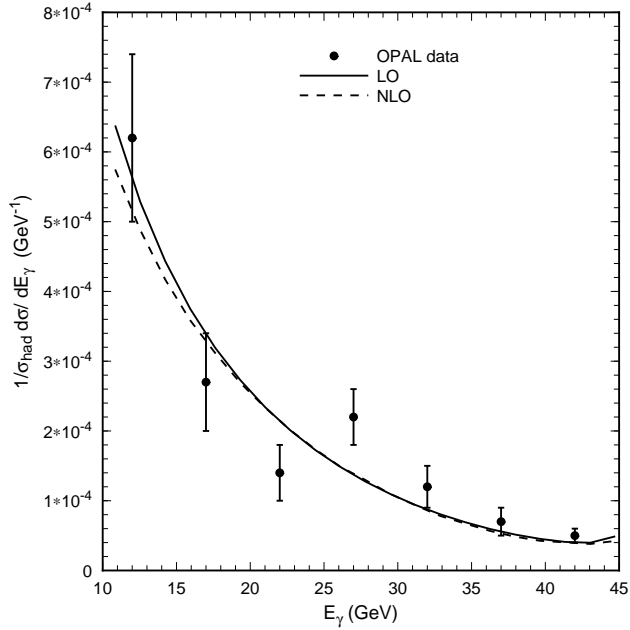,width=9cm}
\caption{The inclusive photon energy distribution 
normalized to the hadronic cross section as measured by the OPAL Collaboration
compared with the ${\cal O}(\alpha)$ and ${\cal O}(\alpha \alpha_s)$
order calculations including the appropriate 
quark-to-photon fragmentation functions 
determined using the ALEPH photon + 1~jet data.}
\label{fig:inclusive}
\end{center}
\end{figure}

The OPAL collaboration has recently measured the inclusive photon distribution
for final state photons with energies between 10 and 42~GeV \cite{OPALinc}.
They have compared their results with the model
estimates of~\cite{grv,owens} and found reasonable agreement 
for $\mu_F \sim M_Z$ in all cases. 
Fig.~\ref{fig:inclusive} shows our (scale independent) predictions 
for the inclusive photon energy distribution at both leading and 
next-to-leading order.
We see good agreement with the data, even though the phase space 
relevant for the OPAL data 
far exceeds that used to determine the fragmentation functions from the ALEPH
photon + 1 jet data \cite{papernew}.

\section{Conclusions}
In summary, we have outlined  the main features 
of the calculation~\cite{big} 
of the `photon' +~1 jet rate at ${\cal O}(\alpha\alpha_s)$. Although 
only next-to-leading order in perturbation theory, this calculation 
contains several ingredients appropriate to the calculation of 
jet observables at next-to-next-to-leading order. In particular, it requires  
to generalize the phase space slicing method of~\cite{gg,kramer}
to take into account contributions where more than one 
theoretically unresolved particle is present in the final state.
The `photon' +~1 jet rate has then been evaluated 
for a democratic clustering algorithm with a Monte Carlo program using 
the hybrid subtraction method of~\cite{eec}.
The results of our calculation, when compared to the data~\cite{aleph}
on the `photon' +~1 jet rate obtained by ALEPH, 
enabled a first determination
of the process independent quark-to-photon fragmentation function
at ${\cal O}(\alpha\alpha_s)$ in a fixed order approach. 
As a first application, we have used this 
function to calculate the `isolated' photon +~1 jet rate in a democratic 
clustering approach at next-to-leading order. The inclusion of the QCD
corrections improves the agreement between theoretical
prediction and experimental data.
Moreover, it was shown that these corrections are moderate,
demonstrating 
the perturbative stability of this particular isolated photon definition.
Finally, we have computed the inclusive photon energy distribution and 
found good agreement with the recent OPAL data.


\begin{thebibliography}{99}


\bibitem{phofrag}
K.~Koller, T.F.~Walsh and P.M.~Zerwas, Z.~Phys. {\bf C2} (1979) 197;\\
E.~Laermann, T.F.~Walsh, I.~Schmitt and P.M.~Zerwas, Nucl.~Phys. {\bf B207}
(1982) 205.

\bibitem{aleph}
ALEPH collaboration: D.~Buskulic et al., Z. Phys. {\bf C69} (1996)
365.

\bibitem{durham} Yu.L.~Dokshitzer, Contribution to the Workshop on Jets 
at LEP and HERA, J. Phys. {\bf G17} (1991) 1441.

\bibitem{andrew}
E.W.N.~Glover and A.G.~Morgan, Z. Phys. {\bf C62} (1994) 311.

\bibitem{grv}
M. Gl\"{u}ck, E. Reya and  A. Vogt, Phys. Rev. {\bf D48} (1993) 116;\\
L.~Bourhis, M.~Fontannaz and J.Ph.~Guillet, Eur. Phys. J. {\bf C2} (1998) 529.

\bibitem{papernew}
A.~Gehrmann--De Ridder and E.W.N.~Glover, preprint DESY-98-068, 
DTP/98/26 
(hep-ph/9806316).


\bibitem{big}
A.~Gehrmann--De Ridder and E.W.N.~Glover,
Nucl. Phys. {\bf B517} (1998) 269.  

\bibitem{curci}
G.~Curci, W.~Furmanski and R.~Petronzio, Nucl. Phys. {\bf B175} (1980) 27;\\
W.~Furmanski and R.~Petronzio, Phys. Lett. {\bf 97B} (1980) 437.

\bibitem{fac}
G.~Altarelli, R.K.~Ellis, G.~Martinelli and S.-Y.~Pi, Nucl. Phys. {\bf B160} 
(1979) 301;\\
P.J.~Rijken and W.L. van Neerven, Nucl. Phys. {\bf B487} (1997) 233.


\bibitem{gg}
W.T.~Giele and E.W.N.~Glover, Phys. Rev. {\bf D46} (1992) 1980.

\bibitem{kramer}
K. Fabricius, I. Schmitt, G. Kramer and G. Schierholz,
Z. Phys. {\bf C11} (1981) 315.


\bibitem{eec} 
E.W.N. Glover and M.R. Sutton, Phys. Lett. {\bf B342} (1995) 375.

\bibitem{letter}
A.~Gehrmann--De Ridder, T. Gehrmann and E.W.N.~Glover, 
Phys. Lett. {\bf B414} (1997) 354. 

\bibitem{OPALinc}
OPAL Collaboration: K. Ackerstaff et al., Eur. Phys. J. {\bf C2} (1998) 39. 


\bibitem{owens}
J.F.~ Owens, Rev. Mod Phys {\bf 59} (1987) 465.



\end{thebibliography}
\end{document}